\documentstyle[12pt,aps,epsf]{revtex}
\vspace{2cm}
\begin{document}
\title{\ \\ \ \\ \ \\ \ \\ \ \\  \ \\
SPECTRAL STATISTICS AT THE ANDERSON TRANSITION: MULTIFRACTALITY OF WAVE 
FUNCTIONS AND THE VIOLATION OF THE NORMALIZATION SUM RULE. }
\author{V.E.Kravtsov}
\address{International Center for Theoretical Physics, P.O. Box 586, 34100 Trieste, Italy\\ and
Landau Institute for Theoretical Physics, Kosygina str. 2, 117940 Moscow, Russia.}
\maketitle
\thispagestyle{empty}
\bigskip
 
{\small  The statistics of energy levels of electrons in a random potential is
considered in the critical energy window near the mobility edge.  
It is shown that
the multifractality of critical wave functions results in the violation of the
normalization sum rule in the thermodynamic limit and leads to the
quasi-Poisson term  $\langle(\delta N)^{2}\rangle=\alpha \langle N\rangle$ 
in the level number variance. The sum rule deficiency $\alpha=\eta/2d$ is 
related to the
multifractality exponent $\eta=d-d^{*}(2)$ in the $d$-dimensional space.}
\bigskip

\section{Introduction}
The energy level statistics in complex quantum systems are the subject of an advance study since the pioneer
works by Wigner, Dyson and Mehta that led to the development of the classical Random Matrix Theory (RMT)
$^{1)}$. The energy level distribution found in the framework of the RMT, known as the Wigner-Dyson (WD) 
statistics, posesses a remarkable property of universality: it depends only on the symmetry of the Hamiltonian
with respect to the time-reversal transformation ${\cal T}$. There are three symmetry classes of Gaussian 
ensembles of random matrices: orthogonal, unitary, and symplectic, labeled by $\beta=1,2$ or $4$, respectively.
Orthogonal and symplectic ensembles correspond to time-reversal invariant Hamiltonians with ${\cal T}^{2}$ equals
$+1$ or $-1$, and the unitary ensemble corresponds to the case of broken time-reversal symmetry.  
These three random matrix ensembles turned out to describe spectral correlations in a great variety of
complex quantum systems, from nuclei to disordered metals, whose classical counterparts exhibit a chaotic 
behavior.

Other universal statistics which apply to describe the spectral correlations in systems with
non-chaotic classical motion, are the well known Poisson statistics. The principal difference between WD and 
Poisson statistics is that there is a phenomenon of level repulsion $^{1)}$ in the former, while the latter
corresponds to completely uncorrelated energy levels.

The system of free electrons in a random potential is a unique example where both universal statistics can
exist depending on the strength of disorder. For low values of impurity concentration $n_{i}<n_{c}$ or
energies $E$ above the mobility edge $E_{c}$, the electron eigenstates are extended and overlap well with each 
other. The spectral correlations in such a metallic phase is well described by WD statistics $^{2,3)}$.
With disorder increasing the system of space dimensionality $d>2$ exhibits the Anderson metal-insulator
transition $^{4)}$ at $n_{i}=n_{c}$ or $E=E_{c}$. Below this transition, in the insulator phase, the typical 
wave function is localized within the volume $\xi^{d}$ which can be centered in an arbitrary point ${\bf r}_{E}$ 
throughout the sample of size $L\gg\xi$. Since two neighbouring in energy eigenstates are typically 
localized in different points in space $|{\bf r}_{E}-{\bf r}_{E'}|\sim L$, there is almost no overlapping 
between them. As a result, the energy levels are almost uncorrelated and obey the Poisson statistics. 

There is, however, a critical region $|E-E_{c}|<\delta E$ near the mobility edge $E_{c}$ where $\xi\propto 
|E-E_{c}|^{-\nu}>L$. The number of eigenstates in this region is equal to $N_{c}=(\delta E)/\Delta\propto 
L^{d-1/\nu}$, 
where $\Delta=(\rho L^{d})^{-1}$ is the mean level spacing, and $\rho$ is the mean density of states. 
According to the well known Harris criterion $\nu d>2$. Therefore the 
number of levels $N_{c}$ can be arbitrary large
in the thermodynamic (TD) limit $L\rightarrow\infty$, and the notion of critical level statistics is 
thus meaningful.

A physical argument that the statistics of eigenvalues is determined by the statistical 
properties of overlapping eigenstates, implies that the critical spectral statistics at the Anderson transition 
might be a representative of the new broad class of universal spectral statistics which is different both 
from WD and from Poisson statistics. The origin for this statement is 
in the special space structure
of critical eigenstates which posess the multifractality $^{5-7)}$. 

Very roughly, one can imagine
such a multifractal structure by considering the space regions $\Omega$ where the modulus of a typical wave
function $|\Psi({\bf r})|$ is bigger than a given number $M$. Then for not too large $M$ the 
space pattern of $\Omega$ will be qualitatively different in the critical spectral region as compared to metal 
and insulator.

In metal, $\Omega$ covers the whole sample volume $L^{d}$ which is "filled" densely by the extended 
wave function.
In the insulator, $\Omega$ is only a small fraction $(\xi/L)^{d}$ of the sample volume that tends to zero in the
TD limit. In contrast to both these cases, a typical pattern of $\Omega$ for a critical eigenstate
is a sparse fractal cluster which is spread throughout the sample but whose total volume is 
proportional to
$L^{d^{*}}\ll L^{d}$. "Multifractality" corresponds to the dependence of the cluster fractal dimension $d^{*}<d$ 
on  the "cut-off level" $M$.

More rigorous definition of the multifractality can be done in terms of the $L$-dependence of 
the moments of inverse 
participation ratio: 
\begin{equation}
\label{IPR}
F_{p}(E)=\sum_{{\bf r}} \langle|\Psi_{n}({\bf r})|^{2p}\delta(E-\varepsilon_{n})
\rangle\propto
\cases{L^{-d(p-1)}; &$metal$;\cr
const.;&$insulator$;\cr
L^{-d^{*}(p) (p-1)}; &$critical$.\cr}
\end{equation}
where $\varepsilon_{n}$ is an exact eigenvalue, $\langle.\rangle$ denotes the ensemble average, and $p$ is an 
integer.

As at large $p$ big slashes of the wave function are enhanced, the dependence of the fractal dimension
$d^{*}(p)$ has the same nature as the dependence $d^{*}(M)$ in the above qualitative picture.

The universality of the critical level statistics first mentioned in $^{8)}$ arises from the scaling
picture of the Anderson transition. The necessary element of any such picture is an existence of
a universal fixed point in the space of relevant parameters which determines all physics in the critical region. 
Within the one parameter scaling by Abrahams, 
Anderson, Licciardello and Ramakrishnan $^{9)}$, the only relevant parameter is the dimensionless conductance
$g(L)$. If the short-scale dimensionless conductance $g(l)$ (where $l$ is the elastic scattering mean free path)
is smaller than the conductance at the fixed point $g_{c}$, the system flows (with increasing $L$) to the 
insulating state, $g(L)\rightarrow 0$, and thus the spectral statistics in the TD limit approaches  Poisson 
statistics. 
If $g(l)>g_{c}$, the system flows to the metal state, $g(L)\rightarrow\infty$, where the spectral statistics is 
WD. However, as long 
as we are in the critical region, the dimensionless conductance is unchanged $g(L)\approx g_{c}$, and thus the 
critical statistics is universal and does not depend on the system size $L$. 

Note that the spectral statistics take a universal form only in the TD 
limit when such parameters as
$l/L$ vanish. Since the mean level spacing $\Delta$ vanishes in this limit too, the meaningful description of 
the level statistics can be done only if all energies are measured in units of $\Delta$, so that separate
levels can be  resolved, no matter how small is $\Delta$.

Physics of critical phenomena which exhibit scaling is described by a number of power-law dependences 
characterizing by the corresponding critical exponents. This is true for the scaling theory of the Anderson 
transition 
too. Within the simple one-parameter scaling $^{9)}$, there is only one non-trivial critical exponent $\nu$
which determines the localization length $\xi\propto|E-E_{c}|^{-\nu}$. All other critical exponents can be 
expressed in terms of $\nu$, the symmetry parameter  $\beta$ and the dimensionality of space $d$. 

In the recent works $^{10,11}$ we
studied analytically the behavior of the correlation function $R(s)$ of two levels separated by an energy 
interval $\omega=s\Delta$ using the simple one-parameter scaling. For this correlation function it is the 
power-law tail $R(s)=-A s^{-\mu}$ at $s\gg 1$ which proves
to depend on properties of the fixed point $g=g_{c}$. The new "spectral" critical exponent $\mu$  
has been related in $^{10,11)}$ with the critical exponent $\nu$ as follows $\mu=1+(\nu d)^{-1}$. 

Multifractality of wave functions is described by
a whole set of additional critical exponents $\eta_{p}=d-d^{*}(p)$ 
which were set  zero in the works $^{10,11)}$. The main 
objective of the present contribution is to show where and how they can show up themselves in the spectral 
statistics.\\ \\ 
 \section{Level number variance and the normalization sum rule.}

The first statement $^{12)}$ on the critical level statistics made in 1988 was about the behavior of the level 
number variance $\Sigma_{2}({\bar N})=\langle(N-{\bar N})^{2} \rangle$ in an energy window of width ${\bar N}\Delta$ 
($1\ll {\bar N}\ll N_{c}$) centered at the mobility edge $E=E_{c}$. This quantity is an integral characteristic
of two-level correlations that can be related $^{1,10)}$ to  
$R(s)$ as follows: 
\begin{equation}
\label{LNV}
\Sigma_{2}({\bar N})=\langle(N-{\bar N})^{2} \rangle=\int_{-{\bar N}}^{+{\bar N}}({\bar N}-|s|)\,R(s)\,ds,
\end{equation}
where the two-level correlation function $R(s)$ is defined in terms of the exact density of states $\rho(E)$:
\begin{equation}
\label{R}
R(s)\rho^{2}=\langle\rho(E+s\Delta)\rho(E) \rangle -\langle\rho(E+s\Delta)\rangle \langle\rho(E)\rangle
\end{equation}
Making use of  the perturbative expression for $R(s)$ derived in $^{13)}$  which is valid in metal for $g(L)\gg 1$, 
and 
then assuming its validity up to  the fixed point $g(L)=g_{c}\sim 1$, it was argued in $^{12)}$ that the level 
number 
variance $\Sigma_{2}({\bar N})=\alpha {\bar N}$ should be linear in ${\bar N}$ with the coefficient $\alpha<1$.
Since then the linear dependence of $\Sigma_{2}({\bar N})$ for ${\bar N}\gg 1$  has been confirmed by a number of 
numerical simulations $^{12, 14, 15}$ on the tight-binding Anderson model.

Thus the $\Sigma_{2}$ statistic at the mobility edge seems to be similar to that in the 
insulator where according to the Poisson law $\Sigma_{2}({\bar N})={\bar N}$, and differs drastically from the
$\Sigma_{2}$ statistic in metal where RMT predicts $^{1)}$ $\Sigma_{2}({\bar N})\propto\ln {\bar N}$.
It is in contrast to other statistics such as the two-level correlation function $^{10,11,16)}$ or spacing 
distribution function $^{8,14,15)}$ which in the critical region behave like $s^{\beta}$ at small $s$ exhibiting the 
level repulsion similar to that in metal.

A challenging issue about the critical level statistics is that two levels repel each other at {\it all}
scales, yet a big number of levels in an energy window fluctuates as if a fraction  of them, $\alpha {\bar N}$, are
completely uncorrelated.

An attempt to understand deeper the nature of the linear term in $\Sigma_{2}$ statistic leads us to re-examening of 
the normalization sum rule. Suppose we have a finite sample with the total number of degrees of freedom
(and the total number of eigenstates) equals $\int \rho(E)\, dE={\cal N}$. 
In contrast to the density of states $\rho(E)$ that takes different values 
for different realizations of disorder,
the number ${\cal N}$ does not fluctuate at all. Using this fact and integrating Eq.(\ref{R}) over $s$ we 
immediately obtain:
\begin{equation}
\label{SR}
\int_{-\infty}^{+\infty}R_{{\cal N}}(s)\,ds\propto
\langle{\cal N}\rho(E)\rangle -\langle{\cal N} \rangle \langle\rho(E) \rangle=0,
\end{equation} 
where the subscript ${\cal N}$ implies that we consider a sample of a 
finite size $L$.

On the other hand, differentiating Eq.(\ref{LNV}) with respect to ${\bar N}$ we have:
\begin{equation}
\label{der}
\frac{d\Sigma_{2}}{d{\bar N}}=\int_{-{\bar N}}^{+{\bar N}}R(s)\,ds.
\end{equation}
Looking at Eqs.(\ref{SR}) and (\ref{der}) one could draw a conclusion $^{10)}$ that $d\Sigma_{2}/d{\bar 
N}\rightarrow 0$ as ${\bar N}\rightarrow\infty$ and thus the linear term in $\Sigma_{2}$ is excluded.
This statement is in fact wrong. The point is that studying the critical level statistics we are bound to
fulfil the inequality ${\bar N}< N_{c}\ll{\cal N}$. Thus at fixed $L$ we cannot take the limit ${\bar N}\rightarrow
\infty$. There is always the remainder integral from $|s|=N_{c}$ to $|s|\sim {\cal N}$ in Eq.(\ref{SR}) where the 
integrand, $R_{{\cal N}}(s)$, is neither critical nor universal. This difficulty could seem to be circumvented if
we take the TD limit as we described before. Then both $N_{c}$ and ${\cal N}$ go 
to infinity, and the remainder
integral should vanish provided that the function $R_{{\cal N}}(s)$ decreases {\it rapidly} enough with increasing 
$s$. It is the last 
condition that makes the situation non-trivial. If the function
$R_{{\cal N}}(s)$ has a small but {\it slowly} decreasing tail $R_{{\cal N}}(s)\sim {\cal N}^{k-1}s^{-k}$ 
($0<k<1$), it becomes possible $^{17,18)}$
that in the TD limit $R_{{\cal N}}(s)$ vanishes on a segment
$|s|\in (N_{c}\rightarrow\infty,{\cal N}\rightarrow\infty)$ 
but the integral of it is not. Moreover, since the function $R_{{\cal N}}(s)$
becomes universal in the TD limit in which $N_{c}/{\cal N}\rightarrow 0$, 
the integral  $\int_{-N_{c}}^{+N_{c}} R_{{\cal N}}(s)\,ds$  
is universal too and must be exactly compensated by the remainder integral in Eq.(\ref{SR}).
This means that the TD limit of the remainder integral is universal $^{17,18)}$ despite the non-universal 
integrand.

One could avoid the above long description, just saying that taking the TD limit and doing the integral in 
Eq.(\ref{SR}) do not necessarily commute. Having an objective to study the {\it universal} spectral statistics
one has to take the TD limit which eliminates the non-universal long tails in $R_{{\cal N}}(s)$. Then it is a 
non-trivial 
question if the normalization sum rule proved rigorously for finite ${\cal N}$ survive taking the TD limit.
It does not if the remainder integral 
tends to  a {\it non-zero} limit $-\alpha$ as $L\rightarrow\infty$. This is just the deficiency of the 
sum rule:
\begin{equation}
\label{DSR}
\int_{-\infty}^{+\infty} R_{\infty}(s)\,ds=\alpha.
\end{equation}
Note that in doing the integral in Eq.(\ref{der}) there is no problem with taking the TD limit, since 
the limits of integration are finite and fixed. Now, using Eq.(\ref{DSR}) and assuming the  power-law tail
in $R_{\infty}(s)=-A s^{-\mu}$ with $2>\mu>1$ we arrive at:
\begin{equation}
\label{var}
\Sigma_{2}({\bar N})=\alpha {\bar N}+\frac{2A {\bar N}^{2-\mu}}{(2-\mu)\,(\mu-1))}+const.
\end{equation}

The second term in Eq.(\ref{var}) is totally determined by the form of the tail in the two-level correlation 
function and reduces to $\ln {\bar N}$  in the RMT limit $\mu\rightarrow 2$. However, the linear in ${\bar N}$
term is due to the violation of the normalization sum rule in the TD limit.\\ \\
\section{The spectral form-factor and the return probability.}

A convenient equivalent formulation of the sum rule violation Eq.(\ref{DSR}) can be done in terms of the
$t\rightarrow 0$ limit of the spectral form-factor 
$2\pi K(t)=\int_{-\infty}^{+\infty}R_{\infty}(s)\,exp[-i s t]\,ds$: 
\begin{equation}
\label{K-O}
\alpha=2\pi K(0).
\end{equation}
The latter can be related to the probability $L^{-d}p(t)$ for a particle to return to the initial point for the 
time $T=t\, \hbar/\Delta$.
A perturbative relationship $^{19)}$ between $K(t)$ and $p(t)$ which is valid in metals for $t\ll 1$ 
( or $T\ll 
\hbar/\Delta$) and is equivalent to the diagrammatic approach $^{13)}$, 
reads: 
\begin{equation}
\label{AIS}
K(t)=(2\pi)^{-2}|t|p(|t|),
\end{equation} 
where $p(t)$ is given by the diffusion propagator $P({\bf q},\omega)$:
\begin{equation}
\label{p-P}
p(t)=\frac{1}{(2\pi \rho)^{2}}\int ds\langle G^{R}_{\varepsilon+s\Delta}({\bf r},{\bf r})\, 
G^{A}_{\varepsilon}({\bf r},{\bf r}) \rangle\,e^{-i s t}=\frac{2}{\beta}\int 
\frac{d\omega}{2\pi}\sum_{{\bf q}} P({\bf q},\omega)\,exp[-i\omega t/\Delta].
\end{equation}
Here $ G^{R(A)}_{\varepsilon}({\bf r},{\bf r'})$ are exact retarded (advanced) Green functions of electrons 
in a random potential, and the factor $2/\beta$ accounts for the number of 'massless' diffusion 
modes for a given symmetry class.

If the multifractality of wave functions is not taken into account, then Eq.(\ref{AIS}) is sufficient
to describe quantitatively the long-range part of two-level correlations in metal and to establish the 
power law tail in $R(s)$ up to the numerical pre-factor  in the critical region. Indeed, in metal we have $ 
P({\bf q},\omega)=(D q^{2}-i\omega)^{-1}$,
where the momentum ${\bf q}$ is quantized $q_{i}=\frac{\pi}{L}n_{i}$ ($n_{i}=0,1,2...$). For sufficiently
small $\omega\ll D/L^{2}$, or a time which is much larger than the diffusion time, the 
diffusing 
wave packet initially $\delta({\bf r})$-shaped at the origin, is spead homogeniously throughout the sample.
It is equivalent  to considering only the zero-mode with ${\bf q}=0$ in the sum, Eq.(\ref{p-P}). Then
one immediately arrives at $p(t)=2/\beta$ and $K(t)=|t|/2\pi^2\beta$. This corresponds to the WD tail
in the two-level correlation function $R(s)= - 1/\pi^2 \beta s^2$. For smaller times $T\ll L^{2}/D$,
the wave packet is spreading diffusively, and $p(t)\propto (Dt)^{-d/2}$. The diffusive propagation of 
a wave packet corresponds $^{19)}$ to
the Altshuler-Shklovskii regime of level correlations $^{13)}$ where $R(s)\propto s^{-(2-d/2)}$.

It turns out $^{10,11)}$ that near the Anderson transition, where the correlation/localization length is 
large $\xi\gg l$, one more regime is possible, which is absent both in a good metal and in a strongly 
localized state. It corresponds to
diffusion at a length scale $r\ll \xi $. According to the one-parameter 
scaling $^{4)}$, diffusion in 
this regime is anomalous with the scale-dependent diffusion coefficient $D(r)=\rho^{-1}r^{2-d}g(r)$ related
to the
the scale dependent conductance $g(r)=g_{c}+\delta g(r)$, where:
\begin{equation}
\label{delta g}
\delta g(r)\sim g_{c}(r/\xi)^{1/\nu}.
\end{equation}
If one sets $g(r)=g_{c}$ as the first approximation, then one arrives at the
 anomalous  diffusion where
$r^{d}\propto t$ is proportional to time for
$T<\hbar/\Delta_{\xi}=\hbar\rho\xi^{d}$. Then the quantity $t p(t)\propto t/r^{d}$ reduces 
to a constant which is irrelevant for the power-law tail in $R(s)$. 
 For $s\gg \Delta_{\xi}/\Delta$ (which corresponds to $t\ll\Delta/\Delta_{\xi}$ )
this tail turns out $^{10,11)}$ to be determined by the 
correction $\delta g(r)$, Eq.(\ref{delta g}):
\begin{equation}
\label{K crit}
K(t)\sim tp(t)=\frac{1}{g(r)}\sim \frac{1}{g_{c}}\left(\frac{g_{c}\Delta_{\xi}t}{\Delta} 
\right)^{\frac{1}{\nu d}}+const.\;\;\;\;\;\; R(s)\sim \frac{1}{g_{c}}\left(\frac{g_{c}\Delta_{\xi}}{\Delta} 
\right)^{\frac{1}{\nu d}}\;s^{-(1+\frac{1}{\nu d})}, 
\end{equation}  
where 
\begin{equation}
\label{D-D}
\frac{\Delta_{\xi}}{\Delta}=\cases{(L/\xi)^{d}; &$l\ll \xi\ll L$;\cr
1; &$\xi\gg L$ \cr}.
\end{equation}
\section{The multifractality of wave functions and  scaling. }
However, the relationship Eq.(\ref{AIS}) does not give even a qualitative description of the critical level 
statistics if multifractality is taken into account. In order to clarify this point let us consider
the representation of $R(s)$ in terms of diagrams with exact diffusion propagators $P({\bf q},\omega)$
and exact vertex parts $\Gamma_{2n+1}(\{{\bf q}_{k}\};\omega)$ which depend on $2n-1$  momenta ${\bf q}_{k}$.
A typical $(2n-1)$-loop diagram is shown in Fig.1. 
\input epsf
\begin{figure}
\epsfysize=4 truecm
\epsfxsize= 6 truecm
\centerline{\epsffile{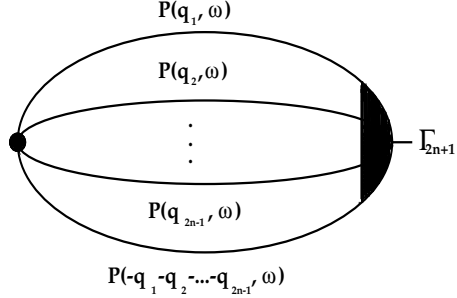}}
\caption[]{A skeleton diagram for $R(s)$.} 
\end{figure}

The relationship between $K(t)$ and $p(t)$ given by
Eq.(\ref{AIS}) exactly corresponds to the one-loop diagram first considered in $^{13)}$.

It is clear that in the critical region all diagrams should be considered. Why then we believe that
the power-law tail is correctly described (up to a numerical pre-factor) by Eq.(\ref{AIS})?
The reason is the scaling behavior of the diffusion propagator. Using the scaling of the diffusion constant
$D(r)\propto r^{2-d}$ one concludes that $Dq^{2}$ scales as $r^{-d}$ or $q^{d}$. 
The scaling dimension of a vertex is equal to zero.
Each loop
in diagrams of Fig.1 adds an extra diffusion propagator that scales like 
$q^{-d}$. This scaling factor is exactly compensated  by a factor 
$q^{d}$ which arises because of an extra
integration over momenta. 
Thus all diagrams shown in Fig.1 have the same scaling dimension and 
should be proportional to the same
power of the quantity $L_{\omega}=(-i\rho\omega)^{-1/d}$ (its absolute value is equal to the mean root 
square of the 
diffusive displacement for the time $1/\omega$) provided that it is the only relevant length in the system in 
the critical region. In case of no multifractality the momentum integration in all diagrams is convergent
at large momenta and the above assumption is correct. However the situation changes if multifractality is 
taken into account. We will see below that in this case all diagrams are divergent, and the inverse upper 
momentum cut-off $\Lambda^{-1}\sim l $ becomes a relevant length too.

Let us introduce the generalized diffusion coefficient $D(q,\omega)$ which obeys the above scaling:
\begin{equation}
\label{D}
D(q,\omega)=g_{c}\rho^{-1}q^{d-2}F(qL_{\omega}),
\end{equation}
where $F(qL_{\omega})$ is a universal scaling function. A usual assumption is that this function
is such that $D(q,\omega)$ is independent of $q$ at $qL_{\omega}\ll 1$ and it is independent of
$\omega$ in the opposite limit. It immediately leads to $D(q,\omega)\propto q^{d-2}$ at large $q$.

However, the multifractality of wave functions should exhibit itself in  the large-$q$ limit of the diffusion 
propagator 
$P({\bf q},\omega)=[q^{2}D(q,\omega)-i\omega]^{-1}$.
Thus it is natural to assume
$^{20)}$ that in case of multifractality $D(q,\omega)\propto q^{d^{*}-2}$ at $qL_{\omega}\gg 1$.
Then Eq.(\ref{D}) leads to:
\begin{equation}
\label{D-m}
q^{2}D(q,\omega)=C g_{c} \rho^{-1} q^{d-\eta} L_{\omega}^{-\eta},
\end{equation} 
where $C$ is a numerical factor, and $\eta=d-d^{*}(2)$. For a diffusion 
propagator that  is expressed in terms of the product 
$|\Psi_{n}|^{2}|\Psi_{m}|^{2}$ (see Eq.(\ref{p-P})), the relevant fractal 
dimension is $d^{*}=d^{*}(2)$.

An immediate consequence of Eq.(\ref{D-m}) is that the return probability given by Eq.(\ref{p-P})
is divergent at large momenta:
\begin{equation}
\label{p-m}
p(t)\sim \frac{(\Lambda L)^{\eta}}{t^{1-\frac{\eta}{d}}}.
\end{equation}
This divergency, in turn, results in the divergency of $K(t)$ in Eq.(\ref{AIS}).
It is instructive to derive how this divergency arises directly from the 
one-loop diagram in Fig.1. In order 
to do that we note that there is a Ward identity that relates the vertex part $\Gamma_{3}({\bf q},\omega)$
with the derivative of the diffusion propagator with respect to $\omega$:
\begin{equation}
\label{WI}
\Gamma_{3}({\bf q},\omega)=P^{-2}({\bf q},\omega)\;\frac{\partial P({\bf q},\omega)}{\partial\omega}=
-\frac{\partial}{\partial\omega}[P^{-1}({\bf q},\omega)].
\end{equation}
Taking into account Eq.(\ref{WI}) and the $\omega$-dependence of the generalized diffusion coefficient at 
$qL_{\omega}\gg 1$ which follows from Eq.(\ref{D-m}), we arrive at:
\begin{equation}
\label{Gamma}
\Gamma_{3}({\bf q},\omega)=\Gamma_{3}^{(0)}+iC\,\frac{\eta}{d}\,g_{c}\,(qL_{\omega})^{d-\eta},
\end{equation}
where $\Gamma_{3}^{(0)}=i$ is a constant.

Eq.(\ref{Gamma}) demonstrates the generic property of  vertex parts
$\Gamma_{2n+1}(\{{\bf q}\},\omega)$ at 
$qL_{\omega}\gg 1$: for $\eta\neq 0$ 
all of them are growing functions of $q$. It is the second term in Eq.(\ref{Gamma}) that leads to the
divergent one-loop diagram for $R(s)$ at $\eta\neq 0$.

One can continue the power-counting for an arbitrary $(2n-1)$-loop diagram using the recursive relation:
\begin{equation}
\label{WI-nn}
\int d^{d}{\bf q}\,\Gamma_{2n+1}(\{{\bf q} \},{\bf q'},{\bf q}';\omega)\,P({\bf q'},\omega)= A_{n}\rho
\Gamma_{2n-1}(\{{\bf q} \},\omega),
\end{equation}
where $A_{n}$ is a numerical coefficient.
Thus we arrive at:
\begin{equation}
\label{G-n}
\frac{\Gamma_{2n+1}}{\Gamma_{2n-1}}\sim [p(s)]^{-1},
\end{equation}
where
\begin{equation}
\label{p-o}
p(s)=\rho^{-1}\int d^{d}{\bf q}\,P({\bf q},\omega)=\frac{\beta}{2}
\int dt \,p(t)\,e^{i s t}
\sim 
(\Lambda L)^{\eta}\,s^{-\frac{\eta}{d}}. 
\end{equation}
Now one can estimate an arbitrary $(2n-1)$-loop diagram for $R(s)$. All of them are divergent
at large momenta so that we consider only the second term in Eq.(\ref{Gamma}) to solve the recursive
relation Eq.(\ref{G-n}). Using also Eqs.(\ref{D-m}) and (\ref{p-o}) we have: 
\begin{equation}
\label{R-2n}
R_{2n-1}(s)\sim \left[\prod_{k=1}^{2n}\int d^{d}{\bf q}_{k}P({\bf q}_{k})
\right]\,\delta({\bf 
q}_{1}+...+{\bf q}_{2n})\Gamma_{2n+1}(\{ 
{\bf q}\};\omega)\sim \frac{1}{s}\,[p(s)]^{n}. 
\end{equation}
The main lesson $^{21)}$ we learn from the above crude power-counting is that as soon as $\eta\neq 0$, the 
power of the divergent parameter $Q=(\Lambda L)^{\eta}$ increases with the number of loops 
$2n-1$ 
in a diagram of Fig.1. Were it not so, there would be no way to get a 
finite TD limit of $R(s)$ as $(\Lambda 
L)\rightarrow\infty$. However, Eq.(\ref{R-2n}) tells us that the possible way out is to assume $^{21)}$ that
the sum $R(s)=\sum_{n=1}^{\infty}R_{2n-1}(s)$ is very much like the geometric progression $Q+Q^{2}+...=
Q/(1-Q)$ which is finite in the limit $Q\rightarrow \infty$. 

Another lesson is that within the power counting the two-level correlation function $R(s)$ can be
represented as a series in $p(s)$ which is the Fourier-transform of the return probability
$p(t)$, all the divergency being absorbed in $p(s)$ or $p(t)$. Thus the possible expression of
$K(t)$ in terms of $p(t)$ that realizes the idea of the 'geometric progression' should be of the form
$K(t)\propto |t|\,p(t)/(1+L[p(t)])$, where $L[p(t)]$ is a {\it linear} in $p(t)$ operator.  

\section{An improved relationship between $K(t)$ and $p(t)$ and the sum rule violation.}
An expression of this kind has been recently derived $^{22)}$ using the idea of parametric diffusion in the
eigenvalue space caused by a random perturbation. It reads:
\begin{equation}
\label{Chalk}
K(t)=\frac{|t|\,p(t)}{(2\pi)^{2}+4\pi \int_{0}^{t}p(t')\,dt'}.
\end{equation}
By the derivation, Eq.(\ref{Chalk}) applies only for small times $T\ll \hbar/\Delta$ ($t\ll 1$).
However, in contrast to Eq.(\ref{AIS}), it gives a qualitatively correct constant limit  
$\lim_{t\rightarrow\infty}K(t)=1/4\pi$, which is yet a factor of two smaller  than the exact 
value $K_{ex}(\infty)=1/2\pi$. 

Now we make use of Eqs.(\ref{Chalk}) and (\ref{p-m}) to relate the deficiency of the sum rule, Eq.(\ref{K-O}) 
with the
multifractality exponent $\eta$. Implying the TD limit and the fact that $p(t)$ is divergent in this limit,
we retain in the denominator of Eq.(\ref{Chalk}) only the term proportional to a divergent parameter $Q=(\Lambda 
L)^{\eta}$. Then we immediately arrive at $^{26)}$:
\begin{equation}
\label{alpha} 
\alpha=2\pi K(0)=\frac{\eta}{2d}=\frac{d-d^{*}(2)}{2d}.
\end{equation}
This expression is a central in our consideration. It gives an answer
to a question {\it why} the normalization sum rule is violated
in the critical region. According to Eq.(\ref{alpha}) it is only because
of the  multifractality of wave functions near the mobility edge.

From the  definition, Eq.(\ref{IPR}), of the fractal dimensions 
$d^{*}(p)$ one can conclude  that the WD statistics in 
metals correspond
formally to $d^{*}(p)\equiv d$ and the Poisson statistics in insulator
correspond to $d^{*}(p)\equiv 0$.
Indeed, in case of WD statistics, the normalization sum rule, Eq.(\ref{SR}), 
is known to survive
the TD limit and no linear term arises in the level number variance, 
Eq.(\ref{var}), as it follows from Eq.(\ref{alpha}) at $d^{*}(2)=d$.  

However, Eq.(\ref{alpha}) is not able to reproduce a correct result
$\alpha=1$ in case of Poisson statistics.
 The reason is that there was an assumption made in the derivation of 
Eq.(\ref{Chalk}) that implies a sort of decoupling between statistics of 
eigenvalues and eigenfunctions. Such an assumption can be
correct only for relatively weak space structure of eigenfunctions, that 
is  for small values of
$\eta$ and away from the strong localization regime.
Thus Eq.(\ref{alpha}) can be considered as the 
first term in the expansion of $\alpha$ in powers of the small
parameter $\eta/d$.

The validity of Eq.(\ref{alpha}) can be checked using the exact results
obtained in 2D systems in the regime of weak localization 
$L\ll \xi=l\,e^{g_{0}}$, where $g_{0}=(2\pi^{2}\rho D)\gg 1$.
It is known $^{5)}$ that in this regime the moments of inverse 
participation ratio behave in the same way as for the critical
eigenstates in Eq.(\ref{IPR}). The corresponding spectrum of
fractal dimensions has been found recently in the framework
of the nonlinear supersymmetric sigma model $^{23)}$ and coincides
with the earlier renormalization group result $^{5,24)}$:
\begin{equation}
\label{spectr}
d^{*}(p)=2-\frac{p}{\beta g_{0}}.
\end{equation}  
On the other hand, as long as the parameter $g_{0}^{-1}$  is small, there is no 
problem to calculate the 
behavior of the two-level correlation function $R_{{\cal N}}(s)$
in the region of large energies $\omega=s\Delta\gg D/L^{2}$
(which are however smaller than the Fermi energy $\varepsilon_{F}\gg \hbar/\tau$).
It has the form $^{25,17)}$:
\begin{equation}
\label{AG}
R_{{\cal N}}(s)=\frac{\delta}{4\pi\beta g_{0}}\,\ln\left[\frac{s^{2}\delta^{2}}{1+s^{2}\delta^{2}} 
\right],
\end{equation}
where the parameter $\delta=(\tau\Delta/\hbar)$ is proportional to ${\cal N}^{-1}$ and vanishes in the 
TD limit.

The expression, Eq.(\ref{AG}) is valid in the ballistic region  
($\omega\sim\hbar/\tau $) too.
It illustrates the existence of a small but slowly decreasing tail which has been discussed   in Section II.
For $s\ll 1/\delta\sim {\cal N}$, the tail is proportional to $\delta\ln (\delta^{-1}/s)$ and is almost 
$s$-independent, while for $s\gg 1/\delta$ it decreases rapidly. This tail vanishes in the TD limit
but the integral of it tends to a finite limit for any fixed lower cut-off $a$:
\begin{equation}
\label{AG-a}
\lim_{{\cal N}\rightarrow \infty}\int_{a}^{\infty} R_{{\cal N}}(s) \,ds= -\alpha=- \frac{1}{2\beta g_{0}}.
\end{equation}
Thus for the 2D system we know (to the leading order in $g_{0}^{-1}$) expressions both for $d^{*}(p)$ and 
$\alpha$ calculated {\it independently}. It is easy to check that the relationship between them 
is really the one that follows from Eq.(\ref{alpha}). 

\section{Discussion and open questions.}
In the above consideration we have shown that the quasi-Poisson linear term in the critical level number 
variance $\Sigma_{2}({\bar N})$ arises because of the  multifractal space structure of wave functions near 
the mobility edge.
Thus we conclude that one of the simplest form all {\it eigenvalue} statistics, the level number variance, 
contains an information
about the fractal dimension $d^{*}(2)=d-\eta$ which characterizes the non-trivial  statistics of {\it 
eigenfunctions}.

It is interesting to compare the value of $\eta$ predicted by Eq.(\ref{alpha})
with that found by the direct numerical investigation $^{27)}$ of statistics of critical eigenfunctions 
in the 3D Anderson model. In order to do that we use the numerically obtained  slope $\alpha$ in the
$\Sigma_{2}({\bar N})$ statistic  at the mobility edge. Simulations on the 3D Anderson model $^{12,14,15)}$
give the value of $\alpha\approx 0.25-0.30$ with the best estimation $^{15)}$ being $\alpha=0.27$.
Then Eq.(\ref{alpha}) predicts $\eta=1.62\pm 0.15$. This is in a surprisingly good agreement 
with an old result $d^{*}(2)=1.7\pm 0.3$ of the direct numerical 
evaluation $^{27)}$ of $d^{*}(2)$.

So far we were interested only in a limit $\alpha=2\pi \lim_{t\rightarrow 0}K(t)$  in 
Eq.(\ref{Chalk}). Within 
the scaling approach we used in Section V, we were able to find only leading power-law dependences 
on $t$. For that reason, the spectral form factor $K(t)$ was independent of $t$ (in the TD limit) 
whatsoever. Small $t$-dependent term in  $2\pi K(t)=\alpha + t^{\mu -1}$ would lead to
a power-law tail in $R(s)\propto s^{-\mu}$ in the same way as in Eq.(\ref{K crit}). A natural question 
which 
arises in this connection is whether the result, Eq.(\ref{K crit}) survives if multifractality and 
the improved relationship
Eq.(\ref{Chalk}) between $K(t)$ and $p(t)$ is taken into account instead of $K(t)\propto |t|\,p(t)$. 

In order to answer this question let us substitute $g_{c}$ for 
$g(r)=g_{c}+\delta g(r)$ 
in Eq.(\ref{D-m}), where $\delta g(r)$ is 
given by Eq.(\ref{delta g}) with $r^{d}\propto t$. The result of such a substitution is the multiplication of $p(t)$ in Eq.(\ref{p-m})
by the same factor of $[1+A(\Delta_{\xi} t/\Delta)^{\frac{1}{\nu d}}]$ as in Eq.(\ref{K crit}).
Now we use Eq.(\ref{Chalk}) to arrive at the result for $K(t)$
that differs only in the numerical coefficient in front of $t^{\frac{1}{\nu d}}$ from 
Eq.(\ref{K crit}). Thus the multifractality and the improved relationship between $K(t)$ and $p(t)$
do leave the result Eq.(\ref{K crit}) unchanged.

However, there could be {\it additional} power-law corrections to Eq.(\ref{p-m}) other than that coming from the correction
to conductance $\delta g(r)$. A possible source of such corrections is the scaling function in Eq.(\ref{D}).
For $L=\xi=\infty$ it depends only on $qL_{\omega}$ but at finite $L$ and $\xi$ it has $1/L$
or $1/\xi$ corrections. In particular for $qL_{\omega}\gg 1$ it should cross over from 
$(qL_{\omega})^{-\eta}$ for $L_{\omega}\ll L<\xi$ to $(qL)^{-\eta}$ for $\xi\gg L_{\omega}\gg L$. 
One can imagine a cross-over of the type 
$F(qL_{\omega},L_{\omega}/L)=(qL_{\omega})^{-\eta}[1+(L_{\omega}/L)^{\zeta}]^{\eta/\zeta}$. It is 
characterized by a new cross-over exponent $\zeta$ which play a role similar to $1/\nu$ in corrections to 
conductance considered above. Thus the question about the exponent $\mu$ in the power-law tail of $R(s)$
is not that simple as it seemed earlier and requires further 
investigations.\\ 
{\bf Acknowledgement} I would like to thank the Institute for Theoretical 
Physics, UCSB for a kind hospitality extended to me during the final 
stage of the work which was supported in part by the NSF Grant 
No. PHY94-07194.

{\bf  Statistique spectrale \`a la transition d'Anderson: 
Multifractalit\'e des fonctions d'ondes et violation d'une 
r\`egle de somme de normalisation.}\\
\\
{\small  La statistique des niveaux d'\'energie d'\'electrons dans 
un potentiel al\'eatoire est consid\'er\'ee dans une fen\^etre 
d'energie critique autours du bord de mobilit\'e. On montre que 
la multifractalit\'e des fonctions d'onde critiques a pour 
cons\'equence la violation d'une r\`egle de somme de normalisation 
dans la limite thermodynamique et donne un contribution quasi-poissonien 
$<(\delta(N)^2> = \alpha <N>$ \`a la variance du nombre de niveaux. 
La deficience $\alpha=\eta/2d$ \`a la r\`egle de somme est reli\'ee 
\`a l'exposant multifractal $\eta=d-d^*(2)$ dans un espace 
d-dimensionnel. }

\end{document}